# Asymptotic homogenization of hygro-thermo-mechanical properties of fibrous networks

E. Bosco[a,b,*], R.H.J. Peerlings[b], M.G.D. Geers[b]

[a]*Materials innovation institute (M2i), P.O. Box 5008, 2600 GA Delft, The Netherlands*
[b]*Department of Mechanical Engineering, Eindhoven University of Technology, P.O. Box 513, 5600 MB Eindhoven, The Netherlands*

**Abstract**

The hygro-thermo-експansive response of fibrous networks involves deformation phenomena at multiple length scales. The moisture or temperature induced expansion of individual fibres is transmitted in the network through the inter-fibre bonds; particularly in the case of anisotropic fibres, this substantially influences the resulting overall deformation. This paper presents a methodology to predict the effective properties of bonded fibrous networks. The distinctive features of the work are i) the focus on the hygro-thermo-mechanical response, whereas in the literature generally only the mechanical properties are addressed; ii) the adoption of asymptotic homogenization to model fibrous networks. Asymptotic homogenization is an efficient and versatile multiscale technique that allows to obtain within a rigorous setting the effective material response, even for complex micro-structural geometries. The fibrous networks considered in this investigation are generated by random deposition of the fibres within a planar region according to an orientation probability density function. Most of the available network descriptions model the fibres essentially as uni-axial elements, thereby not explicitly considering the role of the bonds. In this paper, the fibres are treated as two dimensional ribbon-like elements; this allows to naturally include the contribution of the bonding regions to the effective expansion. The efficacy of the proposed study is illustrated by investigating the effective response for several network realizations, incorporating the influence of different micro-scale parameters, such as fibre hygro-thermo-elastic properties, orientation, geometry, areal coverage.

*Keywords:* fibrous network, asymptotic homogenization, hygro-thermo-expansion,

---

[*]Corresponding author. Tel.: +31-40-247-5169
 *Email address:* `e.bosco@tue.nl` (E. Bosco)

# 1. Introduction

Fibrous networks are encountered in many different engineering applications, ranging from bio-inspired, paper-like to non-woven materials. These materials have an intrinsic multi-scale nature: the constitutive behaviour of the single fibres, the mutual fibres interactions as well as their geometrical features and those of the network all contribute to the effective material response. This work is motivated by paper-based fibrous networks, which are characterized by a strong sensitivity to environmental conditions, such as moisture variations, triggering coupling between their mechanical and hygroscopic response. This is caused by the fact that the hydrophilic paper fibres exhibit large deformations upon humidity changes -up to 20 %. The highly anisotropic swelling of individual fibres and the competition between hygro-mechanical properties taking place in the bonding areas strongly affect the effective paper response (Niskanen, 1998; Larsson and Wagberg, 2008). Focus is here put on the hygro-mechanical properties of paper networks. Note, however, that the approach proposed in this work is general as it allows to establish a scale transition between the level of the underlying micro-structure and the effective hygro-mechanical *and* thermo-mechanical (possibly also hygro-thermo-mechanical) properties of a wider class of fibrous materials. This constitutes a significant step forward towards the development of predictive multi-scale models that are indispensable for many applications.

Extensive work has been done in the literature on the prediction of the effective response of fibrous materials. Analytical models (Cox, 1952; Astrom et al., 1994; Wu and Dzenis, 2005; Tsarouchas and Markaki, 2011) are often based on the assumption of affine deformation of the network, i.e. local strains are equal to the macroscopic deformation. They provide closed form solutions of the effective properties as a function of several micro-structural parameters, for instance the fibre and network geometry, distribution, orientation. These models provide direct insight on the effective behaviour of fibrous materials. However, the approximation of affine deformation is not always realistic, as illustrated e.g. in Hatami-Marbini and Picu (2009). For this reason, more recently, several numerical or computational models of fibrous networks have been developed (Wilhelm and Frey, 2003; Bronkhorst, 2003; Hagglund and Isaksson, 2008;

Hatami-Marbini and Picu, 2009; Kulachenko and Uesaka, 2012; Shahsavari and Picu, 2013; Lu et al., 2014; Dirrenberger et al., 2014). Most of these models are based on a two dimensional description of the material, in which the individual fibres are modelled as a series of trusses or beams with generally isotropic constitutive properties. The fibres are connected to each other at the nodes through inter-fibre bonds, whose role is often not studied explicitly. The main focus of all of these descriptions is on the network's mechanical behaviour, possibly including fracture and damage. Despite the relevance of hygro (thermal) effects on fibrous networks, to the best of our knowledge, the literature lacks contributions specifically dedicated to the analysis of their hygro(thermo)-mechanical response.

The objective of the present study is to investigate the effective hygro-mechanical properties of paper-like fibrous materials based on the analysis of the underlying network, using an asymptotic homogenization approach. Asymptotic homogenization (Sanchez Palencia, 1980; Bakhvalov and Panasenko, 1989; Guedes and Kikuchi, 1990) is a multi-scale technique to model heterogeneous materials with a periodic microstructure. The solution of the equilibrium problem is written as an asymptotic expansion. Inserting this expression into the equilibrium equation leads to an explicit dependence of the displacement (or strain or stress) field on the macro-scale and the micro-scale contributions. This allows to obtain closed form relations for the effective material properties, which are resolved numerically in relation to a given micro-structural geometry.

The micro-structural domain is represented here by a periodic repetition of a network of fibres. Note that the condition of periodicity of the micro-structure is generally not satisfied in real fibrous materials. This issue is circumvented by assuming as a unit-cell a representative volume element, which is large enough to provide sufficient micro-structural information to be representative of a real, disordered micro-scale domain (Drugan and Willis, 1996). The network is generated by depositing the fibres in random positions in a two dimensional region with an anisotropic orientation distribution, see for instance Dodson (1971), Sampson (2009). Contrary to what is generally done in the literature, a key feature of this work is that fibres are modelled as two dimensional ribbon-like elements with an anisotropic hygro-elastic constitutive response. This allows to properly describe the role of the bonding regions in which the coupling between the hygroscopic and mechanical properties of fibres strongly influences the effective hygro-mechanical response (Bosco et al., 2015a,b). The minimum micro-scale cell size, relative to the fibre length, necessary for the convergence of the effective properties is identified by analysing several network realizations. The resulting effective hygro-mechanical properties are investigated by studying the effects of different micro-structural characteristics, in particular the orientation distribution and the coverage, i.e. the ratio between the total area of the fibres and the cell area. Moreover, the adopted asymptotic homogenization approach enables to reconstruct all micro-structural fields, providing insight in the local deformation and displacement due to the interaction between hygroscopic and mechanical phenomena.

This paper is organized as follows. In Section 2, a review on the asymptotic homogenization method is presented, focussing on the calculation of the effective hygro-thermo-mechanical material properties. The network generation and its constitutive characterization are introduced in Section 3. The adopted computational strategy is outlined in Section 4. The results of the study are illustrated in Section 5, in terms of the effective hygro-thermo-mechanical properties as well as the corresponding micro-structural fields. Conclusions are finally given in Section 6.

The following notations for Cartesian tensors and tensor products are used: $a, \mathbf{a}$, $\mathbf{A}$, and $^n\mathbf{A}$ denote, respectively, a scalar, a vector, a second-order tensor, and an $n$th-order tensor. Vector and tensor operations are defined as follows (employing Einstein's summation convention): the dyadic product $\mathbf{ab} = a_i b_j \mathbf{e}_i \mathbf{e}_j$, and the inner products $\mathbf{A} \cdot \mathbf{b} = A_{ij} b_j \mathbf{e}_i$, $\mathbf{A} \cdot \mathbf{B} = A_{ij} B_{jk} \mathbf{e}_i \mathbf{e}_k$, $\mathbf{A} : \mathbf{B} = A_{ij} B_{ji}$, with $\mathbf{e}_i, (i = x, y, z)$ the unit vectors of a Cartesian vector basis. Tensors and tensor operations are represented in a matrix form through Voigt notation: a column and a matrix of scalars are indicated by $\underset{\sim}{a}$ and $\underline{A}$, respectively. The matrix multiplication is defined as $(\underline{A}\underset{\sim}{b})_i = A_{ij} b_j$. Symbol $\boldsymbol{\nabla}$ indicates the gradient operator: $\boldsymbol{\nabla} f = \partial f / \partial x \, \mathbf{e}_x + \partial f / \partial y \, \mathbf{e}_y + \partial f / \partial z \, \mathbf{e}_z$.

## 2. Asymptotic homogenization method

Consider a two dimensional domain $\Omega \in \mathbb{R}^2$, composed by the periodic repetition of a heterogeneous unit-cell of volume $Q \in \mathbb{R}^2$, as shown in Figure 1. The following principles are formulated with reference to Sanchez Palencia (1980); Bakhvalov and Panasenko (1989); Guedes and Kikuchi (1990). Denoting with $H$ and $\eta$ the characteristic sizes of the macroscopic domain and of the micro-structural unit-cell, respectively, it is assumed that a strong separation between scales exists, i.e. $\eta \ll H$. The field quantities governing the elastic problem (e.g. displacement, strain, stress) can thus be considered to vary smoothly at the macroscopic scale, whereas they are periodic at the micro-scale. For this reason, it can be assumed that all quantities explicitly depend on two variables: a slow (macroscopic) variable $\mathbf{X}$ and a fast (microscopic) variable $\mathbf{x} = \mathbf{X}/\eta$.

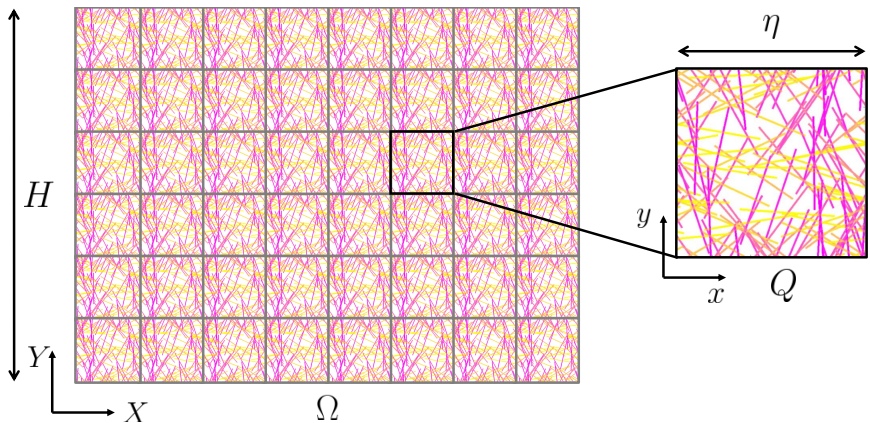

Figure 1: Macroscopic domain and underlying periodic micro-structure.

In the presence of macroscopic mechanical loads and moisture variations, and in the absence of body forces, the equilibrium equation reads

$$\boldsymbol{\nabla} \cdot \boldsymbol{\sigma} = \mathbf{0} \qquad (1)$$

where $\boldsymbol{\sigma}$ is the Cauchy stress tensor, related to the gradient of the displacement field $\mathbf{u}$ through the constitutive relation

$$\boldsymbol{\sigma}(\mathbf{X}) = {}^4\mathbf{C}(\mathbf{x}) : (\boldsymbol{\nabla}\mathbf{u} - \boldsymbol{\beta}(\mathbf{x})\chi(\mathbf{X})) \qquad (2)$$

with ${}^4\mathbf{C}(\mathbf{x})$ the fourth order elasticity tensor and $\boldsymbol{\beta}(\mathbf{x})$ the second order hygro-expansion tensor, related to the moisture variation $\chi$. They are periodic functions in the fast variable $\mathbf{x}$. Note that the thermo-mechanical problem obeys essentially the same equations, in which $\boldsymbol{\beta}(\mathbf{x})$ and $\chi(\mathbf{X})$ are replaced by the thermal-expansion tensor $\boldsymbol{\alpha}(\mathbf{x})$ and by the temperature variation $T(\mathbf{X})$, respectively.

Asymptotic homogenization assumes that the dependence of the solution of (1) on the macroscopic and microscopic levels can be expressed by means of an asymptotic expansion in terms of $\eta$, as

$$\mathbf{u}(\mathbf{X}) = \mathbf{u}^0(\mathbf{X}, \mathbf{X}/\eta) + \eta \mathbf{u}^1(\mathbf{X}, \mathbf{X}/\eta) + \eta^2 \mathbf{u}^2(\mathbf{X}, \mathbf{X}/\eta) + ... \tag{3}$$

The gradient of (3) reads

$$\boldsymbol{\nabla}\mathbf{u} = (\boldsymbol{\nabla}_X \mathbf{u}^0 + \frac{1}{\eta}\boldsymbol{\nabla}_x \mathbf{u}^0) + \eta(\boldsymbol{\nabla}_X \mathbf{u}^1 + \frac{1}{\eta}\boldsymbol{\nabla}_x \mathbf{u}^1) + \eta^2(\boldsymbol{\nabla}_X \mathbf{u}^2 + \frac{1}{\eta}\boldsymbol{\nabla}_x \mathbf{u}^2) + ... \tag{4}$$

where $\boldsymbol{\nabla}_X$ and $\boldsymbol{\nabla}_x$ denote the gradient operators with respect to the slow variable $\mathbf{X}$ and to the fast variable $\mathbf{x}$, respectively. Substituting equation (4) into (2), collecting terms of equal order in $\eta$ and considering the expression up to the first order leads to

$$\boldsymbol{\sigma}(\mathbf{X}) = \frac{1}{\eta}\boldsymbol{\sigma}^{-1}(\mathbf{X}, \mathbf{X}/\eta) + \boldsymbol{\sigma}^0(\mathbf{X}, \mathbf{X}/\eta) + \eta\boldsymbol{\sigma}^1(\mathbf{X}, \mathbf{X}/\eta) + ... \tag{5}$$

with the following definitions

$$\boldsymbol{\sigma}^{-1}(\mathbf{X}, \mathbf{X}/\eta) = {}^4\mathbf{C}(\mathbf{x}) : (\boldsymbol{\nabla}_x \mathbf{u}^0) \tag{6}$$

$$\boldsymbol{\sigma}^0(\mathbf{X}, \mathbf{X}/\eta) = {}^4\mathbf{C}(\mathbf{x}) : (\boldsymbol{\nabla}_X \mathbf{u}^0 + \boldsymbol{\nabla}_x \mathbf{u}^1 - \boldsymbol{\beta}(\mathbf{x})\chi(\mathbf{X})) \tag{7}$$

$$\boldsymbol{\sigma}^1(\mathbf{X}, \mathbf{X}/\eta) = {}^4\mathbf{C}(\mathbf{x}) : (\boldsymbol{\nabla}_X \mathbf{u}^1 + \boldsymbol{\nabla}_x \mathbf{u}^2) \tag{8}$$

The expansion of the stress (5) is finally inserted in (1), obtaining the following expression for the equilibrium equation

$$\frac{1}{\eta^2}\boldsymbol{\nabla}_x \cdot \boldsymbol{\sigma}^{-1} + \frac{1}{\eta}(\boldsymbol{\nabla}_X \cdot \boldsymbol{\sigma}^{-1} + \boldsymbol{\nabla}_x \cdot \boldsymbol{\sigma}^0) + (\boldsymbol{\nabla}_X \cdot \boldsymbol{\sigma}^0 + \boldsymbol{\nabla}_x \cdot \boldsymbol{\sigma}^1) + ... = \mathbf{0} \tag{9}$$

Equation (9) is now required to hold for any order of $\eta$ and for independent variables $\mathbf{X}$ and $\mathbf{x}$. Consider now separately the problem of each order.

2.1. Order minus two problem

The problem of order minus two requires the first contribution of relation (9) to vanish:

$$\boldsymbol{\nabla}_x \cdot ({}^4\mathbf{C}(\mathbf{x}) : (\boldsymbol{\nabla}_x \mathbf{u}^0)) = \mathbf{0} \tag{10}$$

Equation (10) can be satisfied if the solution $\mathbf{u}^0$ is a function that depends only on the slow variable $\mathbf{X}$, i.e.

$$\mathbf{u}^0(\mathbf{X}, \mathbf{X}/\eta) = \mathbf{v}^0(\mathbf{X}) \tag{11}$$

## 2.2. Order minus one problem

Enforcing the second term of (9) to be equal to zero, leads to

$$\nabla_X \cdot (^4\mathbf{C}(\mathbf{x}) : \nabla_x \mathbf{u}^0) + \nabla_x \cdot (^4\mathbf{C}(\mathbf{x}) : (\nabla_X \mathbf{u}^0 + \nabla_x \mathbf{u}^1 - \boldsymbol{\beta}(\mathbf{x})\chi(\mathbf{X}))) = \mathbf{0} \quad (12)$$

Inserting relation (11), the first term of (12) vanishes. The general solution of (12) can be written as the sum of a contribution $\mathbf{v}^1(\mathbf{X})$ depending only on the slow variable, plus two contributions of the fast variable that are linearly related to $\nabla \mathbf{v}^0$ and $\chi$:

$$\mathbf{u}^1(\mathbf{X}, \mathbf{x}) = \mathbf{v}^1(\mathbf{X}) + {}^3\mathbf{N}^1(\mathbf{x}) : \nabla \mathbf{v}^0(\mathbf{X}) + \mathbf{b}^1(\mathbf{x})\chi(\mathbf{X}) \quad (13)$$

where the subscript $\mathbf{X}$ has been dropped from the gradient operator $\nabla$ as $\mathbf{v}^0$ depends on the slow variable only and therefore there is no ambiguity in the definition of its gradient. The influence functions ${}^3\mathbf{N}^1(\mathbf{x})$ and $\mathbf{b}^1(\mathbf{x})$ are the periodic solutions of two boundary value problems, obtained by substituting (13) into (12), defined entirely on the unit-cell and thus depending only on the fast variable $\mathbf{x}$,

$$\nabla_x \cdot (^4\mathbf{C}(\mathbf{x}) : (\nabla_x {}^3\mathbf{N}^1(\mathbf{x}) + {}^4\mathbf{I}^S)) = {}^3\mathbf{0} \quad (14)$$

$$\nabla_x \cdot (^4\mathbf{C}(\mathbf{x}) : (\nabla_x \mathbf{b}^1(\mathbf{x}) - \boldsymbol{\beta}(\mathbf{x}))) = \mathbf{0} \quad (15)$$

where ${}^4\mathbf{I}^S$ is the fourth-order symmetric identity tensor, defined as $I^S_{ijkl} = (\delta_{il}\delta_{jk} + \delta_{ik}\delta_{jl})/2$. For the uniqueness of the solution, in addition to the requirement of periodicity, the following conditions are enforced:

$$\frac{1}{|Q|} \int_Q {}^3\mathbf{N}^1(\mathbf{x}) \mathrm{d}Q = {}^3\mathbf{0} \quad (16)$$

$$\frac{1}{|Q|} \int_Q \mathbf{b}^1(\mathbf{x}) \mathrm{d}Q = \mathbf{0} \quad (17)$$

The functions ${}^3\mathbf{N}^1(\mathbf{x})$ and $\mathbf{b}^1(\mathbf{x})$ describe the periodic micro-fluctuations of the displacement field associated to the macroscopic contributions $\nabla \mathbf{v}_0$ and $\chi$.

## 2.3. Order zero problem

Requiring the third term of (9) to vanish gives

$$\begin{aligned}&\nabla_X \cdot [^4\mathbf{C}(\mathbf{x}) : (\nabla \mathbf{v}^0 + \nabla_x \mathbf{u}^1 - \boldsymbol{\beta}(\mathbf{x})\chi(\mathbf{X}))] + \\ &\nabla_x \cdot [^4\mathbf{C}(\mathbf{x}) : (\nabla_X \mathbf{u}^1 + \nabla_x \mathbf{u}^2)] = \mathbf{0}\end{aligned} \quad (18)$$

In order to obtain the effective hygro-mechanical properties, is it convenient to average (18) over the unit-cell, which by virtue of the divergence theorem and periodicity results in

$$\boldsymbol{\nabla}_X \cdot \bar{\boldsymbol{\sigma}}_0 = \mathbf{0} \tag{19}$$

with

$$\bar{\boldsymbol{\sigma}}^0 = \frac{1}{|Q|}\int_Q {}^4\mathbf{C}(\mathbf{x}):(\boldsymbol{\nabla}\mathbf{v}^0 + \boldsymbol{\nabla}_x\mathbf{u}^1 - \boldsymbol{\beta}(\mathbf{x})\chi(\mathbf{X}))\mathrm{d}Q = {}^4\overline{\mathbf{C}}:(\boldsymbol{\nabla}\mathbf{v}^0 - \overline{\boldsymbol{\beta}}\chi) \tag{20}$$

The final equality is obtained by inserting expression (13) into (20), leading to the following definitions of the effective elastic tensor ${}^4\overline{\mathbf{C}}$ and of the effective hygro-expansion tensor $\overline{\boldsymbol{\beta}}$:

$$^4\overline{\mathbf{C}} = \frac{1}{|Q|}\int_Q {}^4\mathbf{C}(\mathbf{x}):(\boldsymbol{\nabla}_x{}^3\mathbf{N}^1(\mathbf{x}) + {}^4\mathbf{I}^S)\mathrm{d}Q \tag{21}$$

$$\overline{\boldsymbol{\beta}} = \frac{1}{|Q|}{}^4\overline{\mathbf{C}}^{-1}:\int_Q {}^4\mathbf{C}(\mathbf{x}):(\boldsymbol{\beta}(\mathbf{x}) - \boldsymbol{\nabla}_x\mathbf{b}^1(\mathbf{x}))\mathrm{d}Q \tag{22}$$

The tensors ${}^4\overline{\mathbf{C}}$ and $\overline{\boldsymbol{\beta}}$ are constant with respect to the fast variable; inserting (20) into (19) yields a problem in terms of the macroscopic variable $\mathbf{X}$ only that provides the macroscopic solution $\mathbf{v}^0$. On the other hand, inserting expressions (11) and (13) into relation (18) allows to calculate the second order contribution $\mathbf{u}^2$. Going to higher orders results in higher order effective media (Boutin, 1996; Peerlings and Fleck, 2004), which may be relevant if the scale separation is not very large. Here, the response to essentially constant hygroscopic loading is considered, for which this is not needed. Furthermore, the solution of the macro-problem for $\mathbf{v}^0$ is trivial under these conditions, once the effective properties ${}^4\overline{\mathbf{C}}$ and $\overline{\boldsymbol{\beta}}$ are known.

Note finally that departing from the thermo-mechanical problem in equations (1)-(2) and following the same procedure, the expression for the effective thermal-expansion coefficient is found as

$$\overline{\boldsymbol{\alpha}} = \frac{1}{|Q|}{}^4\overline{\mathbf{C}}^{-1}:\int_Q {}^4\mathbf{C}(\mathbf{x}):(\boldsymbol{\alpha}(\mathbf{x}) - \boldsymbol{\nabla}_x\mathbf{a}^1(\mathbf{x}))\mathrm{d}Q \tag{23}$$

with $\mathbf{a}^1$ solution of the cell problem

$$\boldsymbol{\nabla}_x \cdot ({}^4\mathbf{C}(\mathbf{x}):(\boldsymbol{\nabla}_x\mathbf{a}^1(\mathbf{x}) - \boldsymbol{\alpha}(\mathbf{x})) = \mathbf{0} \tag{24}$$

where periodicity and vanishing fluctuations $\mathbf{a}^1$ over $Q$, similarly to (17), are enforced. If in (2) the thermal and the hygroscopic contributions are present at the same time (as e.g. in Bacigalupo et al. (2016)), relations (21),(22) and (23) should be simultaneously used and represent the effective properties of the hygro-thermo-mechanical problem.

## 3. Fibrous network model

*3.1. Geometrical features*

The unit-cell describes a periodic two dimensional network of fibres characterized by constant length $l$ and width $w$, and rectangular cross section $w \times t$, where $t$ is the thickness of the fibre. The mean areal coverage $\bar{c}$ is defined as the ratio between the total area occupied by the fibres and the area of the unit-cell, $\bar{c} = A_f/Q$. The coverage can be interpreted as the average number of fibres through-the-thickness that can be found in the unit-cell and thus is a measure of the average thickness of the network. The number of fibres $n_f$ within a network of given coverage is thus obtained as

$$n_f = \frac{\bar{c}\,Q}{lw} \qquad (25)$$

The networks used in this paper have been generated departing from a uniform random point field. The center of each fibre is located in the square unit-cell $Q = [0, L] \times [0, L]$, with $L = 1$. To each point an angle $\theta$ is associated, according to a wrapped Cauchy orientation distribution (Cox, 1952):

$$f(\theta) = \frac{1}{\pi}\frac{1-q^2}{1+q^2-2q\cos(2\theta)} \qquad (26)$$

where $-\pi/2 < \theta \leq \pi/2$ is the angle between the fibre axis and the horizontal direction and $0 \leq q < 1$ is a measure of the anisotropy of the network orientation.

Periodicity of the cell is achieved by trimming fibre segments that exit the cell boundaries and periodically copying them to the opposite cell edges. An example of a network of fibres with $l = L/2$, aspect ratio $l/w = 20$, coverage $\bar{c} = 0.25$ and a uniform fibre orientation distribution($q = 0$) is illustrated in Figure 2(left).

*3.2. Fibre and bond constitutive models*

The constitutive response of a single fibre can be described through transversely isotropic material properties, with respect to a local coordinate system $(\ell, t, z)$ defined along its principal directions, see for instance Bergander and Salmén (2002); Borodulina et al. (2015). The general constitutive law (2) is specified for the fibre's transversely isotropic elasticity tensor $^4\mathbf{C}^f$ and the hygro-expansion tensor $\boldsymbol{\beta}^f$. In Voigt notation, they can be expressed as

$$\underline{C}^f = \begin{pmatrix} \frac{E_\ell}{(1-\nu_{\ell t}\nu_{t\ell})} & \frac{\nu_{t\ell}E_\ell}{(1-\nu_{\ell t}\nu_{t\ell})} & 0 \\ \frac{\nu_{\ell t}E_t}{(1-\nu_{\ell t}\nu_{t\ell})} & \frac{E_t}{(1-\nu_{\ell t}\nu_{t\ell})} & 0 \\ 0 & 0 & G_{\ell t} \end{pmatrix} ; \quad \underset{\sim}{\beta}^f = \begin{pmatrix} \beta_\ell \\ \beta_t \\ 0 \end{pmatrix} \qquad (27)$$

Here, $E_\ell$ and $E_t$ are the logitudinal and transverse moduli of elasticity, respectively, while $G_{\ell t}$ is the in-plane shear modulus. The in-plane Poisson's ratios are indicated with $\nu_{\ell t}$ and $\nu_{t\ell}$, with $E_\ell \nu_{t\ell} = E_t \nu_{\ell t}$. The constants $\beta_\ell$ and $\beta_t$ indicate the longitudinal and transverse hygro-expansion coefficients of the fibre, respectively.

The constitutive quantities (27) are referred to a fibre oriented at an angle $\theta^{(k)}$. For the purpose of the subsequent derivations, they have to be expressed with respect to the global reference system $(x, y, z)$ according to Roylance (1996):

$$\underline{C}^{(k)} = \left[\underline{R}\ (\underline{A}^{(k)})^{-1}\underline{R}^{-1}(\underline{C}^f)^{-1}\underline{A}^{(k)}\right]^{-1} \qquad (28)$$

$$\underset{\sim}{\beta}^{(k)} = \underline{R}\ (\underline{A}^{(k)})^{-1}\underline{R}^{-1}\underset{\sim}{\beta}^f \qquad (29)$$

where, having defined $c = \cos\theta^{(k)}$ and $s = \sin\theta^{(k)}$, the transformation matrix $\underline{A}^{(k)}$ and the Reuters matrix $\underline{R}$ read

$$\underline{A}^{(k)} = \begin{pmatrix} c^2 & s^2 & 2sc \\ s^2 & c^2 & -2sc \\ -sc & sc & c^2 - s^2 \end{pmatrix} \quad \underline{R} = \begin{pmatrix} 1 & 0 & 0 \\ 0 & 1 & 0 \\ 0 & 0 & 2 \end{pmatrix} \qquad (30)$$

Bonds are identified as the regions where two or more fibres overlap. Full kinematic compatibility between the fibres composing a bond is assumed. The bonds elasticity tensor $^4\mathbf{C}^b$ and the hygro-expansion tensor $\boldsymbol{\beta}^b$ are thus calculated by using a Voigt average (Bosco et al., 2015b):

$$^4\mathbf{C}^b = \sum_{k=1}^{n_b} \lambda^{(k)} {}^4\mathbf{C}^{(k)} \qquad (31)$$

$$\boldsymbol{\beta}^b = (^4\mathbf{C}^b)^{-1} : \sum_{k=1}^{n_b} \left(\lambda^{(k)} {}^4\mathbf{C}^{(k)} : \boldsymbol{\beta}^{(k)}\right) \qquad (32)$$

where $n_b$ is the number of fibres composing the considered bond and $\lambda^{(k)} = t/(t\bar{c})$ is the relative thickness of a fibre with respect to the average thickness of the network.

Note finally that, as the model is two dimensional, the porosity of the system in the thickness direction is in principle not captured. To correctly describe the freestanding, unbonded fibre segments as the the weaker portions of the network, their stiffness should be also normalized with respect to the average network thickness. The correct input value to be used in the homogenization procedure is ${}^4\mathbf{C}^{f(k)*} = {}^4\mathbf{C}^{f(k)}/\bar{c}$. This essentially corresponds to relation (31),the presence of a voided region in the thickness direction is considered.

## 4. Computational strategy

### 4.1. Network discretization

The fibrous network developed in Section 3 is discretized using two dimensional elements. First, the unit-cell domain is subdivided into a regular grid of $n_e$ square (finite) elements of edge $l_e = w/\xi$; here $\xi$ is an integer, with $\xi \geq 1$. An element $e$ is considered to belong to a given fibre if its geometrical center lies inside the area of the fibre, of dimensions $l \times w$. This results in a discrete approximation of the fibre boundaries by zigzag edges. A detail of the finite element mesh of the network illustrated in Figure 2(left) is shown in Figure 2(right). In this case, on average, five elements across the width of the fibre have been used, i.e. $\xi = 5$. Whereas the local stress and strain concentrations may be affected by the resulting zigzag geometry, it has been verified that if $\xi$ is sufficiently large (e.g. $\xi = 5$) this influences only minimally the resulting macroscopic properties, which are the main quantities of interest for the present investigation.

The discretized network generated according to the above procedure, completed with the corresponding constitutive properties given in Section 3.2 and the requirements of periodicity and zero average, is used as the input for the solution of the unit-cell problems (14) and (15).

### 4.2. Finite element solution

The functions ${}^3\mathbf{N}^1(\mathbf{x})$ and $\mathbf{b}^1(\mathbf{x})$ are determined by solving the cell-problems (14) and (15) numerically. Their corresponding weak forms can be approximated by finite

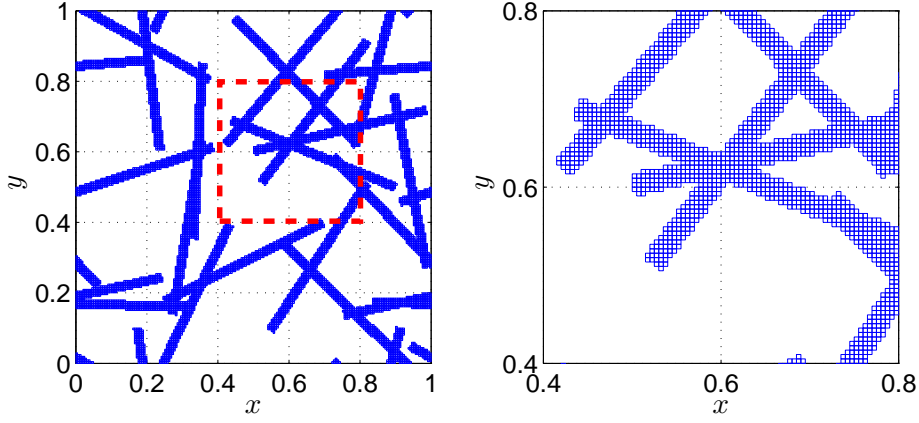

Figure 2: Example of a network configuration with coverage $\bar{c} = 0.25$ (left) and detail showing the finite element discretization (right).

element interpolation as

$$\underline{\mathbf{K}} \cdot {}^3\underset{\sim}{\mathbf{N}}^1 = {}^3\underset{\sim}{\mathbf{F}} \tag{33}$$

$$\underline{\mathbf{K}} \cdot \underset{\sim}{\mathbf{b}}^1 = \underset{\sim}{\mathbf{f}} \tag{34}$$

where the columns ${}^3\underset{\sim}{\mathbf{N}}^1$ and $\underset{\sim}{\mathbf{b}}^1$ contain the nodal values of ${}^3\mathbf{N}^1$ and $\mathbf{b}^1$, and the following matrices/columns are given by

$$\underline{\mathbf{K}} = \int_{Q^h} \boldsymbol{\nabla} \underset{\sim}{N} \cdot {}^4\mathbf{C}(\mathbf{x}) \cdot \boldsymbol{\nabla} \underset{\sim}{N}^T \mathrm{d}Q \tag{35}$$

$$ {}^3\underset{\sim}{\mathbf{F}} = -\int_{Q^h} \boldsymbol{\nabla} \underset{\sim}{N} \cdot ({}^4\mathbf{C}(\mathbf{x}) : {}^4\mathbf{I}^S) \mathrm{d}Q \tag{36}$$

$$\underset{\sim}{\mathbf{f}} = \int_{Q^h} \boldsymbol{\nabla} \underset{\sim}{N} \cdot ({}^4\mathbf{C}(\mathbf{x}) : \boldsymbol{\beta}(\mathbf{x})) \mathrm{d}Q \tag{37}$$

Here, $Q^h$ denotes the discretized network geometry as discussed in Section 4.1, while $\boldsymbol{\nabla} \underset{\sim}{N}$ is the column containing the partial derivatives of the interpolation functions. More details on the finite element procedure can be found in Peerlings and Fleck (2004) and Pinho-da-Cruz et al. (2009).

Periodic boundary conditions are enforced

$$ {}^3\underset{\sim}{\mathbf{N}}^{1L} = {}^3\underset{\sim}{\mathbf{N}}^{1R} \quad \text{and} \quad {}^3\underset{\sim}{\mathbf{N}}^{1T} = {}^3\underset{\sim}{\mathbf{N}}^{1B} \tag{38}$$

$$\underset{\sim}{\mathbf{b}}^{1L} = \underset{\sim}{\mathbf{b}}^{1R} \quad \text{and} \quad \underset{\sim}{\mathbf{b}}^{1T} = \underset{\sim}{\mathbf{b}}^{1B} \tag{39}$$

where superscripts "$L - R$" and "$T - B$" denote corresponding points on opposite boundaries: left and right and top and bottom, respectively.

The effective hygro-mechanical properties given by relations (21) and (22) can thus be computed by using the same finite element interpolation functions as

$$^4\overline{\mathbf{C}} = \frac{1}{|Q^h|}\int_{Q^h} {}^4\mathbf{C}(\mathbf{x}) : (\boldsymbol{\nabla}\underset{\sim}{N}^T\, {}^3\underset{\sim}{\mathbf{N}}^1 + {}^4\mathbf{I}^S)\mathrm{d}Q \tag{40}$$

$$\overline{\boldsymbol{\beta}} = \frac{1}{|Q^h|}{}^4\overline{\mathbf{C}}^{-1} : \int_{Q^h} {}^4\mathbf{C}(\mathbf{x}) : (\boldsymbol{\beta}(\mathbf{x}) - \boldsymbol{\nabla}\underset{\sim}{N}^T\underset{\sim}{\mathbf{b}}^1)\mathrm{d}Q \tag{41}$$

Bilinear quadrilateral finite elements have been used. The algorithm for the network generation and discretization and the finite element solution of the cell-problems have been implemented in MATLAB.

## 5. Results

### 5.1. Network parameters

Several network configurations have been considered by assuming a set of different coverages $\bar{c} = [0.25, 0.5, 1, 2, 5, 10]$. The unit-cell edge has been defined as $L = 1$. In order to evaluate the minimum unit-cell size necessary to obtain convergence of the effective properties, different ratios between the cell edge $L$ and the fibre length $l$ have been examined: $L = [0.5, 1, 2, 4, 8]\, l$. The fibre width $w$ has been assumed as $w = l/50$. Note that the value of the fibre thickness $t$ is not a parameter for the model, as the local constitutive properties are entirely defined by the ratio between the thickness of fibres and of the bonds, relative to the network average thickness. The size of the finite element edge is obtained by choosing $\xi = w/l_e = 5$. The transverse elastic modulus $E_t$ and the shear modulus $G_{\ell t}$ are defined with respect to the longitudinal stiffness $E_\ell = 1$ through the constants $\alpha = E_\ell/E_t = 6$ and $\gamma = E_\ell/G_{\ell t} = 10$. The Poisson's ratio $\nu_{\ell t}$ is taken as $\nu_{\ell t} = 0.3$, while $\nu_{t\ell} = \nu_{\ell t}/\alpha = 0.05$. The hygro-expansion coefficient in the transverse direction is expressed in terms of the longitudinal one as $\beta_t = 20\beta_\ell$, with $\beta_\ell = 1$. The chosen ratios for the longitudinal and transverse properties are typical values of the hygro-mechanical properties of pulp fibres (Niskanen, 1998; Schulgasser and Page, 1988).

### 5.2. Isotropic case: uniform fibre orientation

Networks characterized by a uniform fibre orientation distribution, i.e. $q = 0$ in (26), are first considered. The cell problems (33) and (34) are solved through the finite element procedure discussed in Section 4.2, providing the influence functions ${}^3\mathbf{N}^1(\mathbf{x})$ and $\mathbf{b}^1(\mathbf{x})$. The effective properties ${}^4\overline{\mathbf{C}}$ and $\overline{\boldsymbol{\beta}}$ are then computed through (40) and (41) for several network realizations. For low coverages ($\bar{c} = [0.25, 0.5]$), ten realizations of the unit cell have been considered; for higher coverages ($\bar{c} = [1, 2, 5, 10]$), five network configurations have been generated.

*5.2.1. Influence of the cell size on the effective properties*

The effect of the ratio between the size of the system, given by the cell edge $L$, and the length of a single fibre $l$ on the effective properties is investigated to define the minimum cell size to be used in the further analyses. Figure 3(left) and Figure 3(right) illustrate the average effective stiffness components $(\overline{C}_{xx} + \overline{C}_{yy})/2E_\ell$ and the average hygro-expansion coefficients $(\overline{\beta}_{xx} + \overline{\beta}_{yy})/2\beta_\ell$, respectively, as a function of the ratio between the cell and the fibre length, $L/l$. The averages of the effective properties between the $x-$ and $y-$ direction have been considered in order to minimize the influence of possible anisotropy due to the randomness of the fibre orientation. Dots refer to the values corresponding to different realizations of the micro-structure, whereas continuous lines represent the average. The different colours indicate different coverages. All curves asymptotically converge to a limit value as $L/l$ increases. This limit value increases with the network coverage $\bar{c}$. Due to the stochastic nature of the generated networks, for given coverage and cell size, the effective stiffness and the effective hygro-expansion vary from realization to realization. In general, the variability from case to case decreases with increasing $L/l$ due to the fact that a larger unit-cell contains more statistical information. It can also be noticed that data related to expansive properties show a larger scatter with respect to the ones related to mechanical properties. This is due the larger contrast in the anisotropy of the hygroscopic properties (compared to mechanical properties) of the single fibre. As a criterion for the choice of the cell size, the effective properties are considered sufficiently converged if their value averaged over the different network realizations is within a range of 5% from the asymptote. Networks with lower coverages present a higher scatter and require a larger cell size ($L \geq 2l$) to converge to the asymptotic value of the considered material property. For higher coverages, e.g. $\bar{c} \geq 5$, the effective material properties approach the asymptote even for cell sizes equal to or smaller than the fibre length $l$. For such high coverages, the network essentially responds as a heterogeneous solid. The cell size has to be larger than any correlation length of the network, either geometrical or in terms of material properties. In Hatami-Marbini and Picu (2009), it has been shown that random networks have a spatial correlation of the order of $l$. Based on these considerations, a minimum cell size of $L = 2l$ is assumed for any coverage. All further results have thus been generated with cells with these dimensions, even though they may be unnecessary large for high coverages.

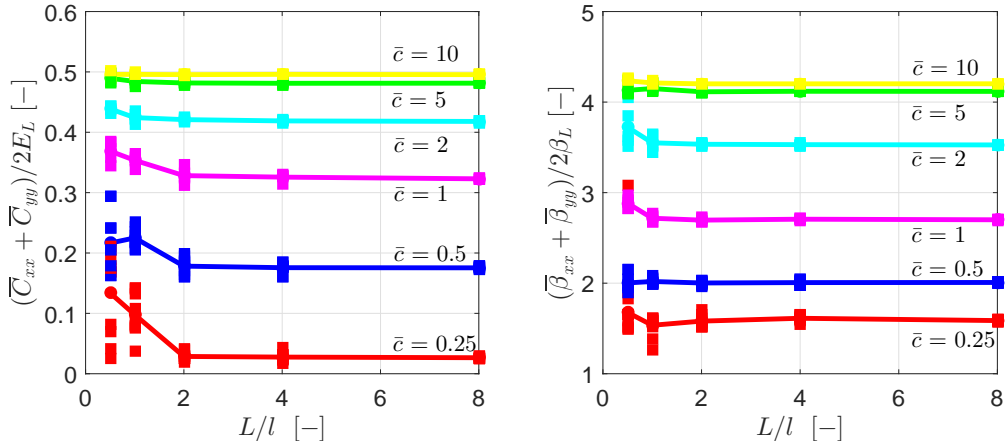

Figure 3: Average ffective stiffness components $(\overline{C}_{xx} + \overline{C}_{yy})/2E_\ell$ (left) and average hygro-expansion components $(\overline{\beta}_{xx} + \overline{\beta}_{yy})/2\beta_\ell$ as a function of cell size relative to fibre length, $L/l$ (right). A minimum cell size of $L = 2l$ is assumed.

*5.2.2. Effective properties as a function of the coverage*

The hygro-mechanical effective properties are shown as a function of the coverage $\bar{c}$ in Figure 4 in terms of the average values of $(\overline{C}_{xx} + \overline{C}_{yy})/2E_\ell$ (left) and $(\overline{\beta}_{xx} + \overline{\beta}_{yy})/2\beta_\ell$ (right) for all considered network realizations. Both the effective stiffness and the hygro-expansion increase with the coverage and approach the Voigt bound for the network with all orientations present, i.e. $\bar{c} \to \infty$ (magenta lines in the Figure), computed as

$$^4\overline{\mathbf{C}}^V = \int_{-\pi/2}^{\pi/2} f(\theta)^4\mathbf{C}^{(k)}(\theta)\,\mathrm{d}\theta \qquad (42)$$

$$\overline{\boldsymbol{\beta}}^V = (^4\overline{\mathbf{C}}^V)^{-1} : \int_{-\pi/2}^{\pi/2} f(\theta)^4\mathbf{C}^{(k)}(\theta) : \boldsymbol{\beta}^{(k)}(\theta)\,\mathrm{d}\theta \qquad (43)$$

where the probability density function (26) for $q = 0$ reads $f(\theta) = 1/\pi$; $^4\mathbf{C}^{(k)}(\theta)$ and $\boldsymbol{\beta}^{(k)}(\theta)$ are functions of the angle $\theta$ according to (28) and (29). The obtained results reveal that denser networks are characterized by a stiffer response and a larger hygro-expansion. Whereas for mechanical properties this behaviour is intuitive, for expansive properties this depends on the interaction between the longitudinal and transverse expansion coefficients and the elastic moduli. The bonding regions, for the parameters chosen in Section 5.1, have a higher resulting expansion than the free fibre segments. For this reason, sparse networks have a hygro-expansion closer to the longitudinal coefficient $\beta_\ell$, $(\overline{\beta}_{xx} + \overline{\beta}_{yy})/2\beta_\ell \approx 1$, i.e. as if the fibres are pin-jointed. The high number of constraints in denser networks yields a larger effective expansion, in the limit (approximatively for $\bar{c} \geq 5$) approaching the Voigt upper bound.

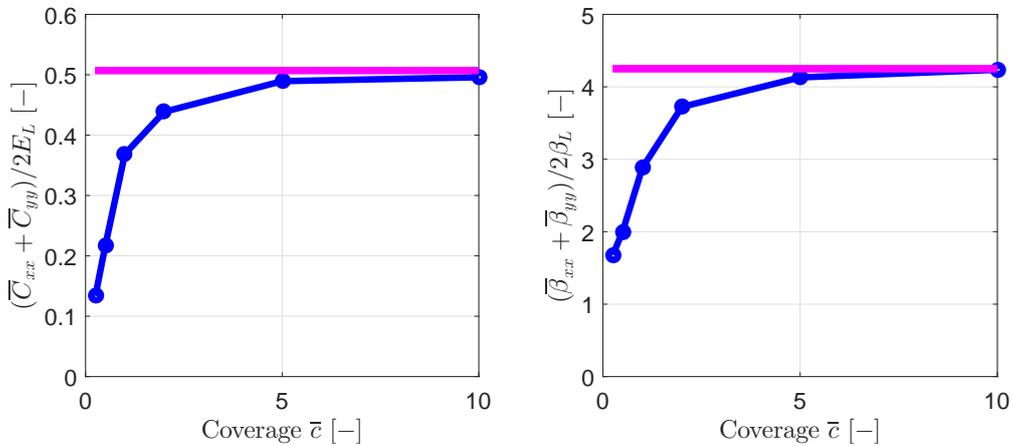

Figure 4: Average effective stiffness components $(\overline{C}_{xx} + \overline{C}_{yy})/2E_\ell$ (left) and average expansion components $(\overline{\beta}_{xx} + \overline{\beta}_{yy})/2\beta_\ell$ (right) as a function of the coverage $\bar{c}$.

*5.2.3. Local fields*

The influence functions ${}^3\mathbf{N}^1(\mathbf{x})$ and $\mathbf{b}^1(\mathbf{x})$ are solutions of the cell problems (33) and (34). Tensor ${}^3\mathbf{N}^1(\mathbf{x})$ describes the micro-fluctuation of the displacement field due to the macroscopic deformation $\boldsymbol{\nabla}\mathbf{v}^0$. Its component $N^1_{xxx}$ is shown in Figure 5(left) for a network of coverage $\bar{c} = 0.25$. It can be interpreted as the micro-structural correction to the overall displacement field in the $x$ direction associated with a unit variation of the $xx$ component of the overall strain $\boldsymbol{\nabla}\mathbf{v}^0$. Vector $\mathbf{b}^1(\mathbf{x})$ illustrates the micro-fluctuation of the displacement field associated to the (macroscopic) moisture variation $\chi$ under constrained expansion conditions, i.e. for $\boldsymbol{\nabla}\mathbf{v}^0 = \mathbf{0}$. This is due to the "true" periodicity requirement for $\mathbf{b}^1(\mathbf{x})$ on the cell boundary, expressed by (39). Vector $\mathbf{b}^1(\mathbf{x})$ thus contains information on both the mechanical and hygro-expansive contributions to the micro-structural displacement field. Figure 5(right) shows the component $b_x^1$. A qualitative comparison with the purely mechanical micro-fluctuation reveals a rougher profile of the expansion related micro-fluctuation with respect to the mechanical one, due to the higher ratio of anisotropy assumed in the fibre expansion properties.

Finally, the purely hygro-expansive micro-fluctuation $\mathbf{b}^{1,free}(\mathbf{x})$ can be computed assuming a free expansion condition, where the macroscopic strain is $\boldsymbol{\nabla}\mathbf{v}^0 = \overline{\boldsymbol{\beta}}\chi$, as follows

$$\mathbf{b}^{1,free}(\mathbf{x}) = {}^3\mathbf{N}^1(\mathbf{x}) : \boldsymbol{\nabla}\mathbf{v}^0 + \mathbf{b}^1(\mathbf{x})\chi = ({}^3\mathbf{N}^1(\mathbf{x}) : \overline{\boldsymbol{\beta}} + \mathbf{b}^1(\mathbf{x}))\chi \qquad (44)$$

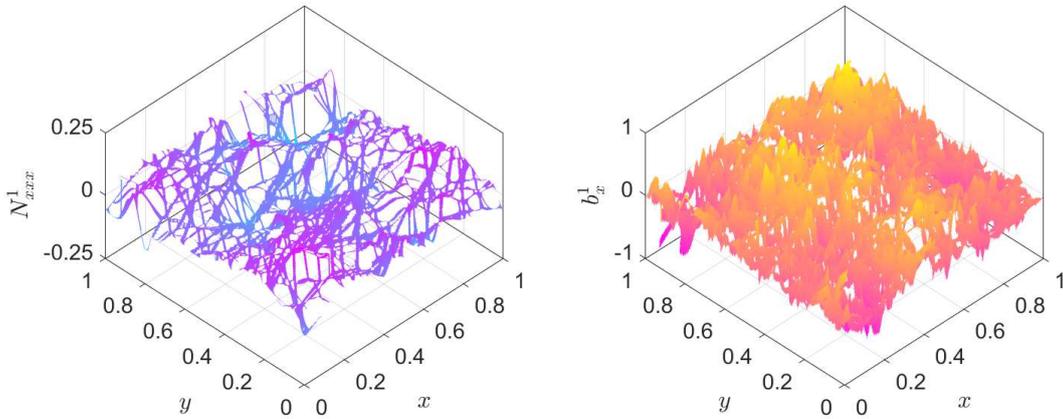

Figure 5: (left) Component $N_{xxx}^1$ of the mechanics related micro-fluctuation field. (right) Expansion related micro-fluctuation field $b_x^1$.

The adopted asymptotic homogenization strategy allows to reconstruct all the local fields at the cell level by combining the obtained macroscopic fields and micro-fluctuation fields. Consider as an example a network subjected to a macroscopic variation of moisture content in an zero average stress state. The macroscopic strain is purely expansive; the macroscopic displacement is thus $\mathbf{v}^0 = \overline{\boldsymbol{\beta}}\cdot\mathbf{x}\chi$. The local displacement field can be calculated by considering the zero and the first order terms in (3), i.e. summing $\mathbf{v}^0$ to the micro-fluctuation $\mathbf{b}^{1,free}(\mathbf{x})$ associated to free expansion given by (44). Figure 6 illustrates the displacement normalized with $\beta_\ell$ due to free expansion for a moisture content variation of $\chi = 0.2$ of two networks of coverages $\bar{c} = 0.25$ and $\bar{c} = 1$. Black and magenta lines show the fibres in the undeformed and deformed configurations, respectively. The elongation of fibres is less than the transverse expansion and less than the expansion of the bonding regions. This explains why, as expected from the obtained overall hygro-expansive coefficients, a high coverage network with a larger number of bonds has a higher resulting expansion than a sparse network.

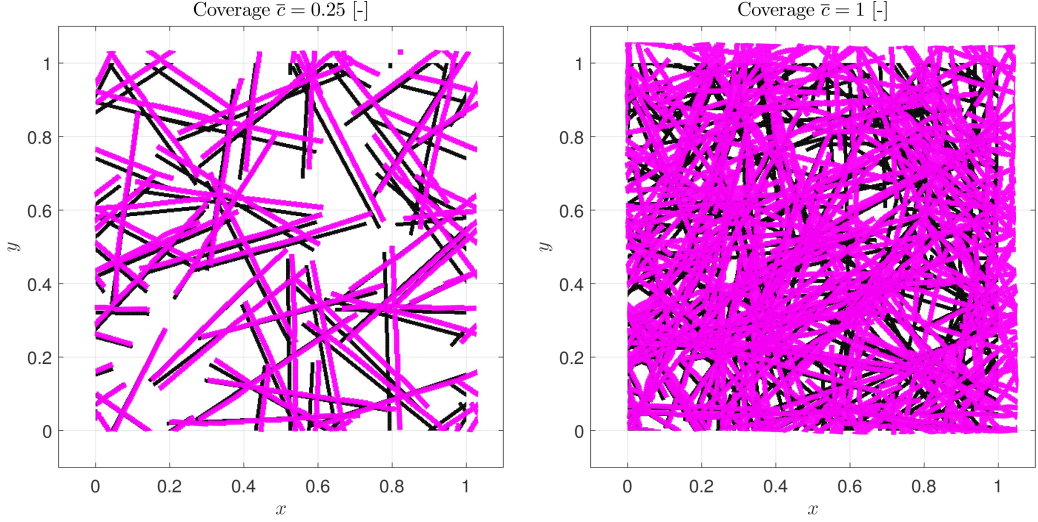

Figure 6: Deformed configuration (magenta lines) $\mathbf{u}/\beta_\ell$ due to free expansion for networks of coverages $\bar{c} = 0.25$ (left) and $\bar{c} = 1$ (right).

The network local strains can be obtained again using (3), specified for instance for the case of free expansion as described above. The local maximum and minimum principal strains, $\varepsilon_{max}$ and $\varepsilon_{min}$, normalized by $\beta_\ell \chi$, are shown in Figure 7 for the network of coverage $\bar{c} = 1$, as in Figure 6(right). In the free fibre segments, the strain distribution is relatively homogeneous as it is mostly due to the expansive contribution only. The interplay between mechanics and expansion caused by the misorientation of the different fibres passing through the bonds creates higher fluctuations in the strain distribution. Consequently, internal stresses arise in these regions, despite the fact that the average stress is zero.

*5.3. Anisotropic case: influence of the orientation distribution*

The influence of the fibre orientation on the effective hygro-mechanical properties is investigated by considering different levels of anisotropy in the orientation distribution function (26), in addition to the isotropic case ($q = 0$): $q = 0.25, q = 0.5$ and $q = 0.75$. Five network realizations have been considered for each analysis.

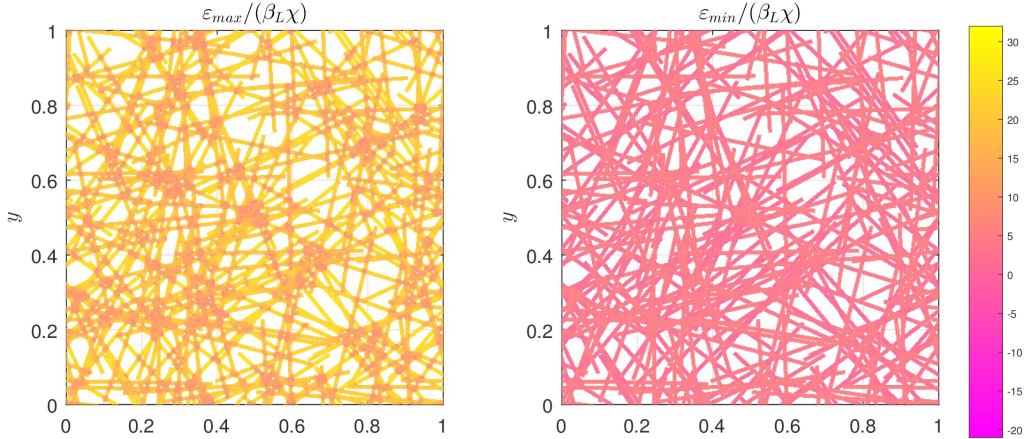

Figure 7: Maximum principal strain $\varepsilon_{max}/\beta_\ell\chi$ (left) and minimum principal strain $\varepsilon_{min}/\beta_\ell\chi$ (right) due to free expansion in a network of coverage $\bar{c} = 1$.

### 5.3.1. Effective properties

Figure 8 presents the elastic stiffness components and the expansion coefficients in $x$ direction (blue lines) and $y$ direction (magenta lines), normalized with respect to the longitudinal elastic modulus $E_\ell$ and the expansion coefficient $\beta_\ell$, respectively. Continuous lines refer to coverage $\bar{c} = 2$, while dashed lines refer to coverage $\bar{c} = 10$. The results have been obtained for a cell size of $L = 2l$, as determined from the analysis of the isotropic fibre orientation distribution. The dots in Figure 8 show that the scatter of the results for the different realizations is limited and that therefore the anisotropy of the fibre orientation does not affect the minimum cell size. Like for the isotropic case, a higher coverage yields higher effective properties. The effective stiffness $\overline{C}_{xx}$ increases for increasing anisotropic fibre distribution, whereas $\overline{C}_{yy}$ decreases. The hygro-expansion coefficients show the opposite trend: $\overline{\beta}_{xx}$ is lower than $\overline{\beta}_{yy}$ and decreases for higher anisotropy, asymptotically approaching the value of $\beta_\ell = 1$. This would occur in the limit for $q \to 1$, representing an ideal network in which the fibres are all aligned along the $x$ direction. On the other hand, $\overline{\beta}_{yy}$ is more influenced by the anisotropy of the orientation, due to both the high contrast between the longitudinal and transverse expansive properties as well as the interaction with the mechanical response. For $q \to 1$, $\overline{\beta}_{yy} \to \beta_t = 20\beta_\ell$. Note that for the lower coverage, $\bar{c} = 2$, at $q = 0$, statistical isotropy is not perfectly achieved and there is a small difference between the material response in both directions.

The effective hygro-mechanical properties of networks with different levels of anisotropy are shown in Figure 9 as a function of the coverage. Continuous lines relate to $q = 0.25$ while dashed-dotted lines relate to $q = 0.5$. Magenta and blue curves refer to properties in the $x$ and $y$ direction, respectively. The corresponding Voigt bounds are calculated through (43) by inserting in $f(\theta)$ the corresponding value of $q$ and are illustrated in Figure 9 by the pink (for $x$ direction properties) and cyan (for $y$ direction properties) horizontal lines parallel to the $x-$axis. The dashed purple curve illustrates the isotropic reference case ($q = 0$) as presented in Figure 4. For any coverage, a preferential orientation of fibres along the $x$ direction leads to increasing mechanical stiffness $\overline{C}_{xx}/E_\ell$ and decreasing $\overline{C}_{yy}/E_\ell$, as well as a decreasing expansion $\overline{\beta}_{xx}/\beta_\ell$ and increasing $\overline{\beta}_{yy}/\beta_\ell$. Moreover, for increasing anisotropy, the expansion coefficient $\overline{\beta}_{xx}$ reaches the Voigt average faster. Finally, it can be observed that the Voigt estimate captures the effective response for high coverages independently of the anisotropy level.

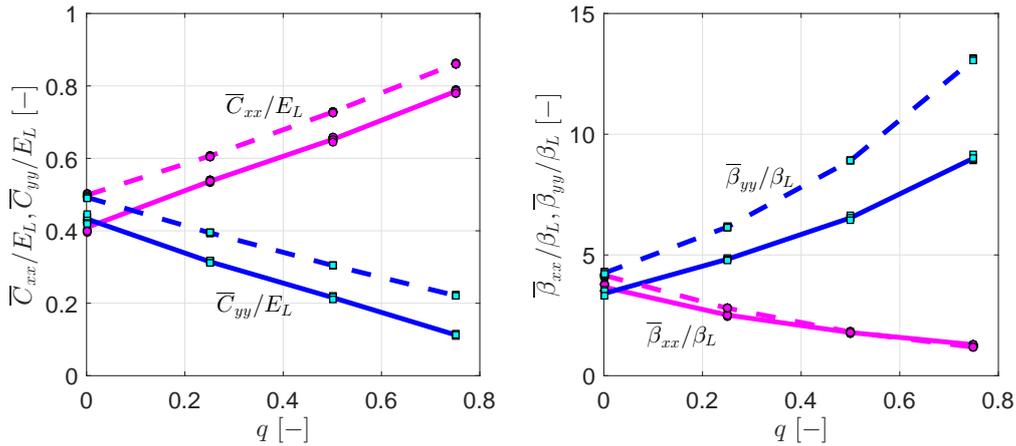

Figure 8: Effective stiffness components $\overline{C}_{xx}/E_\ell$ and $\overline{C}_{xx}/E_\ell$ (left) and effective expansion $\overline{\beta}_{xx}/\beta_\ell$ and $\overline{\beta}_{xx}/\beta_\ell$ (right), as a function of the anisotropy of the fibre orientation $q$, for coverages $\bar{c} = 2$ (continuous lines) and $\bar{c} = 10$ (dashed lines).

*5.3.2. Local fields*

Local fields are studied for networks with different levels of anisotropy. Here, two networks of $\bar{c} = 0.25$ are considered: the first for an orientation distribution with $q = 0$ (i.e. the same shown in Figure 6(left)), the second with $q = 0.5$. They are subjected to a macroscopic moisture content variation of $\chi = 0.4$. The local displacement field (normalized with respect to $\beta_\ell$) is calculated according to the procedure illustrated in Section 5.2.3 and is shown in Figure 10. A strong influence of the fibre orientation can be seen in the network expansion, revealing a larger expansion in the $y$ direction as the anisotropy of the network increases. This depends on the larger number of bonding regions in the $y$ direction compared to the $x$ direction, which are characterized by a higher expansion than the unbonded fibre segments. This observation is consistent with the trend shown in Figure 9(right), although related a to different coverage.

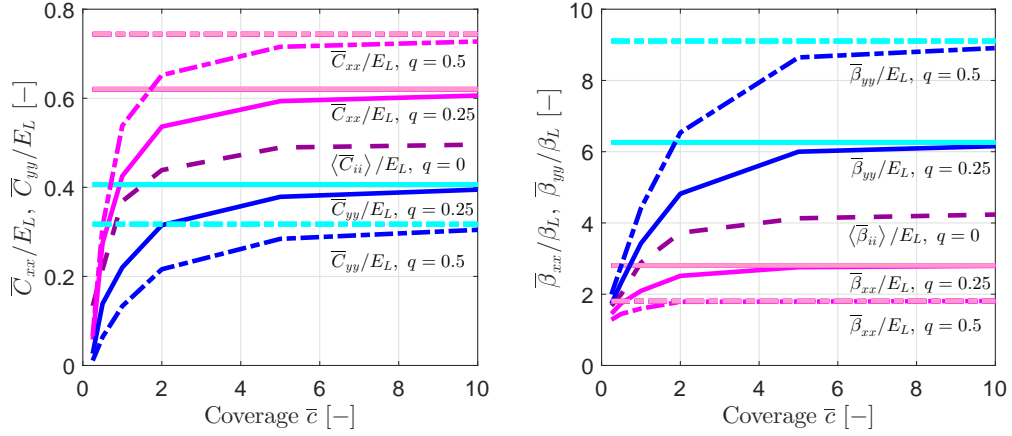

Figure 9: Effective stiffness components $\overline{C}_{xx}/E_\ell$, $\overline{C}_{yy}/E_\ell$ (left) and expansion components $\overline{\beta}_{xx}/\beta_\ell$, $\overline{\beta}_{xx}/\beta_\ell$ (right) as a function of the coverage $\bar{c}$, for different network orientation distributions.

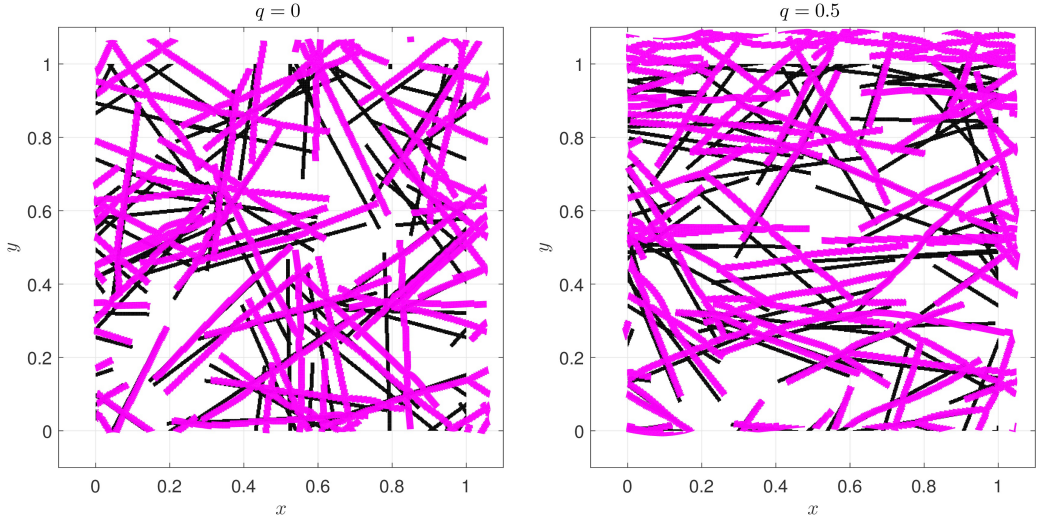

Figure 10: Effect of the anisotropic orientation on the network deformation. Displacement $\mathbf{u}/\beta_\ell$ (magenta lines) due to free expansion for networks of coverages $\bar{c} = 0.25$, with $q = 0$ (left) and $q = 0.5$ (right).

## 6. Conclusions

This paper presented a study of the effective hygro-thermo-elastic response of bonded fibrous materials based on the analysis of the underlying network of fibres. The approach has been applied to the hygro-mechanical response of paper fibrous networks, whereby it can be used without any loss of generality for analysing the effective thermo-mechanical response of a large class of fibrous materials (e.g. non-woven, cellular materials). Asymptotic homogenization has been applied to bridge the micro-scale material properties to the effective macro-scale moduli. This is done by calculating local fluctuation fields through the solution of two boundary value problems, one related to mechanical and one to hygro-expansive response, at the network level. Based on these micro-fluctuations the effective properties are computed. The fibrous network has been modelled as a periodic arrangement of two dimensional unit-cells, in which the fibres are randomly distributed and oriented according to a given orientation distribution. Fibres have been represented as two dimensional ribbon-like elements with transversely isotropic hygro-elastic properties. This type of description enables to naturally capture the interaction between mechanical and expansive phenomena in the bonding areas and its influence on the effective material response. The major outcomes of this contribution are:

- The minimum cell size $L$ that provides converged effective hygro-mechanical properties for any value of coverage has been established to be two times the length of the fibre $l$. For high coverages, for which the material substantially behaves as a heterogeneous solid, the size can be even equal or lower than $l$.

- The effective stiffness and expansion have been estimated as a function of the areal coverage. They increase with increasing coverage, approaching for approximatively $\bar{c} \geq 5$ a limit value that corresponds with the Voigt bound.

- The influence of the anisotropic orientation distribution on the effective properties has been investigated. For the case study of paper fibrous networks, for increasing orientation the stiffness increases in the preferential direction of alignment of the fibres, while it decreases in the orthogonal direction. An inverse trend characterizes the effective hygro-expansion. The Voigt average captures the effective properties at high coverages independently from the anisotropy of the orientation.

- The local network fields (local strains, network deformation) can be reconstructed through the micro-fluctuations provided by the solution of the cell problems.

The comparison of the theoretical results with experimental data at the two scales would certainly strengthen the validity of the model. This would also provide useful insight in the material characterization at the fibre and network scale.

Finally, the asymptotic expansion that describes the displacement field has here been truncated at the first order. Higher order contributions can incorporate scale effects in relation to the fibre length scale. Their influence on the effective hygro-thermo-mechanical response of fibrous materials will be the subject of forthcoming work.

## Acknowledgements

This research was carried out under project number M61.2.12458 in the framework of the Research Program of the Materials innovation institute M2i (www.m2i.nl).